# Robust linear magnetoresistance in WTe$_2$

Xing-Chen Pan[1], Yiming Pan[1], Juan Jiang[2], Huakun Zuo[3], Huimei Liu[1], Xuliang Chen[4], Zhongxia Wei[1], Shuai Zhang[1], Zhihe Wang[1], Xiangang Wan[1], Zhaorong Yang[4], Donglai Feng[2], Zhengcai Xia[3], Liang Li[3], Fengqi Song[1,*], Baigeng Wang[1,*], Yuheng Zhang[4], Guanghou Wang[1]

[1]National Laboratory of Solid State Microstructures, Collaborative Innovation Center of Advanced Microstructures, and College of Physics, Nanjing University, Nanjing, 210093, P. R. China

[2]State Key Laboratory of Surface Physics, Collaborative Innovation Center of Advanced Microstructures, Department of Physics, and Advanced Materials Laboratory, Fudan University, Shanghai 200433, China

[3]Wuhan National High Magnetic Field Center, Huazhong University of Science and Technology, Wuhan 430074, China

[4]High Magnetic Field Laboratory, Chinese Academy of Sciences, Hefei, Anhui 230031, China.

---

[*] Corresponding authors. F.S. (songfengqi@nju.edu.cn) and B.W. (bgwang@nju.edu.cn). Fax:+86-25-83595535

**Abstract**

Unsaturated magnetoresistance (MR) has been reported in WTe$_2$, and remains irrepressible up to very high field. Intense optimization of the crystalline quality causes a squarely-increasing MR, as interpreted by perfect compensation of opposite carriers. Herein we report our observation of linear MR (LMR) in WTe$_2$ crystals, the onset of which is first identified by constructing the mobility spectra of the MR at low fields. The LMR further intensifies and predominates at fields higher than 20 Tesla while the parabolic MR gradually decays. The LMR remains unsaturated up to a high field of 60 Tesla and persists, even at a high pressure of 6.2 GPa. Assisted by density functional theory calculations and detailed mobility spectra, we find the LMR to be robust against the applications of high field, broken carrier balance, and mobility suppression. Angle-resolved photoemission spectroscopy reveals a unique quasilinear energy dispersion near the Fermi level. Our results suggest that the robust LMR is the low bound of the unsaturated MR in WTe$_2$.

**Introduction**

Following the extensive studies on graphene, broad interest in new kind of two-dimensional atomic-layer materials has been quick to emerge due to a desire for advanced new devices based on Van der Waal bonded heterostructures[1, 2]. Highly efficient gate-tunable photocurrent generation [3], field-effect tunneling transistors [4] and light-emitting diodes [5] have respectively been demonstrated in vertical graphene heterostructures, graphene–$MoS_2$–graphene stacks and even more complex but carefully designed structures with BN/graphene/$WS_2$. The versatile functionality is well accommodated in the engineered nanodevices, leading to efforts of seeking out candidate materials with multiple physical properties.

$WTe_2$ consists of heavy elements and is the rare semi-metal among all the materials contained in the roadmap of A. Geim [1]. A new physics has recently been discovered, relating to its extremely high (of order $10^6$ times) magnetoresistance (MR), which remains unsaturated up to high fields of 60 Tesla [6]. The angle resolved photoemission spectroscopy (ARPES) soon has visualized its equal sizes of the electron/hole pockets at the Fermi level at low temperatures [7], as also described in terms of quantum oscillations [8, 9]. Additionally, this crystal has superconducting properties when a pressure of 2.5 GPa is applied [9, 10] in which it well reproduces the dome-like pressure-superconductivity temperature phase diagram of $MoS_2$ [11]. One-dimensional transport through its atomic chains is suggested, based on longitudinal MR measurements under optimized configurations [10]. Most recently, an ARPES study using polarized light demonstrated a circular dichroism, indicating a new physics related to exotic spin textures [12].

Here we demonstrate the introduction of a robust and unsaturated linear MR (LMR) for this crystal, which we found to survive against a high field, broken carrier balance, and mobility suppression. This is first visualized using the mobility spectra of the MR at low fields, and it remains unsaturated up to a high field of 60 Tesla. It supplies a strong support to the MR unsaturation and universality. Our work implies that $WTe_2$ is a multifunctional candidate for advanced vertical heterostructures.

**The signature of the LMR: a mobility spectrum analysis**

An elaborate transport description of the single crystal of $WTe_2$ can be achieved using a mobility spectrum method, which extracts the carrier density, mobility, and strength of all the effective transport components (see Supplementary Information). Such spectra have been useful in successfully obtaining transport descriptions of a number of multi-band systems, e.g., a two-dimensional electron gas [13], doped semiconductors [14], semi-metals [15] and iron-based superconductors [16, 17]. Herein, it is used to demonstrate the signature of LMR, as well as the balance of electron- and hole-like carriers at low temperatures, as expected.

We collected magneto-transport data for $WTe_2$ single crystal at various temperatures. The crystalline structure of $WTe_2$ is shown in **Figure 1(a)**. The inset of **Figure 1(b)** shows the high-resolution transmission electron microscopic image of the single crystal and its Fourier transformation. A photograph of a typical sample is shown in the upper panel of **Figure 1(c)**, where its smooth, shiny appearance confirms its crystalline nature. A magnetic field (H) was applied parallel to the c-axis and a current was passed along the a-axis (W chains) respectively. The MR was defined by $(\rho_{xx}(H)-\rho(0))/\rho(0)$. The MR and Hall resistivity from 1.7 to 77 K are shown for a typical sample in **Figure 1(c)**. The MR increases in an almost parabolic way (according to $H^2$) in $WTe_2$, reproducing the results shown in previous reports [6] [10, 18]. A strong nonlinearity in the magnetic field dependence of the Hall resistivity can be seen, suggesting a multiple-band picture of the transport in $WTe_2$. Interestingly, a "turn on" effect in the longitudinal resistivity $\rho_{xx}$[6] may be observed in the temperature dependence of the Hall coefficient (defined as $\rho_{yx}/\mu_0 H$) for various magnetic fields.

The mobility spectra were constructed according to the MR data. In a material with a complicated Fermi surface consisting of multiple Fermi pockets such as $WTe_2$[8, 12, 18], it is a real challenge to assign the preliminary number of effective bands using the conventional approach to construct a multi-band model. The mobility spectrum can be used to address this problem, as shown in **Figure 2**. We first obtained the normalized conductivity matrix $\sigma_{ij}(H)/\sigma(0)$ from the resistivity matrix $\rho_{ij}(H)$ at 4.2 K as shown in **Figure 2(a)**. To produce an analytical representation of the transport data, we fitted a normalized conductivity matrix using a linear combination of the Lorentzian components. The Kronig-Kramer (KK) transformations were then performed on the analytical representations in order to separate the contributions made by the electron- and hole-like carriers. The partial conductivities of all the electron- and hole-like components obtained from the KK transformations are shown in the green and violet curves in **Figure 2(a)**, respectively. The contributions of the

electron- and hole-like carriers can be deconvolved over the temperature range between 1.7 and 35 K [19]. We finally obtained the mobility spectra at various temperatures by calculating the normalized conductivity matrix of the electron- and hole-like carriers as shown in **Figure 2(b)**. Although WTe$_2$ has a complicated Fermi surface with nine pockets[12], a single effective carrier component was clearly identified for the electron and hole sides at all temperatures considered.

The unsaturated MR in WTe$_2$ is believed to be a property of a compensated semi-metal, which occurs because the exactness of the compensation leads to a cancellation of the Hall-induced electrical field[6]. Our mobility spectra provide experimental evidence on the balance between the electron- and hole-like carriers in WTe$_2$. To obtain the carrier density at various temperatures, we calculated the integrals of the mobility spectra for the data from 1.7 to 35 K, and the results are shown in **Figure 2(c)**. The carrier mobility at various temperatures can also be calculated from the positions of the peaks in the mobility spectra (**Figure 2(d)**). The mobilities remain fairly steady below 10 K but gradually decrease when the temperature increases above 20 K. The carrier concentrations of the electron- (n) and hole- (p) like carriers are compared in the framework of (n-p)/(n+p) as shown in **Figure 2(e)**, where the two carriers are found to be in balance within an uncertainty of 2 percent. This supports theoretical predictions that an exact compensation (n=p) at low temperatures contributes to the non-saturating parabolic MR up to 60 Tesla in this material [6, 7, 20].

Interestingly, there is a long tail that shows an extreme mobility in the spectra at

4.2 K, while the spectra are rather more localized at 35 K, as shown in **Figure 2(b)**. These long tails in the mobility spectra can best be understood in terms of the total conductivity tensor having a nontrivial component ~ 1/H, which does not exist in the semi-classical two-band isotropic model (with a parabolic dispersion). This has previously been related to the LMR in $BaFe_2As_2$ and $FeSe$[16, 17]. We therefore believe this to be the signature of the LMR onset in the low-field transport of our material.

**LMR developed at up to 60T and under high pressure**

LMR is seen again and indeed predominates over parabolic MR at temperatures above 20 T, and remains unsaturated up to 60 T. As shown in **Figure 2(e)**, the MR data were measured at different temperatures for a pulsed high field. We found that all the MR data continued to increase irresistibly while the field increased up to an extreme value of 60 T, although some of the characteristics were due to Shubnikov de Haas oscillations (SdHOs) at lower temperatures. We also note that the LMR was further developed at higher fields as seen in the MR at 4.2 K, where the MR maintained its parabolic (semiclassical) behavior at low fields and became entirely linear above 20 T. This is clearer in the $H^2$ coordinate, as shown in **Figure 2(f)**. A linear curve is seen for the data for 100 K in **Figure 2(f)**, which implies a parabolic MR at a temperature of 100 K because the semi-classical component still dominates the total conductivity tensor at this temperature. The blue curve clearly deviates from the semiclassical MR in the high-field range at 4.2 K. From all the transport data, it is clear that the LMR trend tends to predominate at higher fields and lower temperatures, indicating its nature as an intrinsic property of the $WTe_2$ crystal.

Our experiments at high pressures confirm the robustness of the LMR. As shown in **Figure 3(a)**, we obtained our MR data at 1.3, 4.3 and 6.2 GPa, and compared our results with data obtained under ambient pressure. The data under high pressure were collected at 5 K, at which pressure-induced superconductivity does not occur[9]. The experiment under ambient pressure was carried out at 4.2 K. Firstly, MR is greatly suppressed by more than one order of magnitude under high pressure as reported previously [8-10]. SdHOs cannot be observed at high pressures, indicating the suppression of the carrier mobility. Secondly, a clear LMR is observed at a pressure of 6.2 GPa. The first magnetic field derivatives of the MR are shown in **Figure 3(b)**. At ambient pressure, dMR/dH is nearly H-linear, indicating an almost $H^2$ MR in the range of fields used. As the pressure increases, dMR/dH becomes nonlinear against H. At 6.2 GPa, dMR/dH becomes constant, and is independent of H. This indicates a perfect and robust LMR at high pressures, while the semiclassical parabolic MR, which is the origin of the vast MR unsaturation, fades out because of suppressed mobility or broken carrier balance.

As previously discussed, no obvious structural transition was found under pressures comparable with those used in our experiments . However, the density of states was very sensitive to the pressure. The density of states increased rapidly as the pressure increased, suggesting that the carrier density might vary greatly [9]. Moreover, SdHOs at high pressures indicated that the Fermi surface was very sensitive to pressure. The results of both the calculations and the experiments strongly suggest that the balance between electron and hole may be broken under high pressure. This implies that LMR becomes dominant while the balance of electron- and hole-like carriers is broken, as required for semiclassical MR unsaturation. In other words, a robust LMR may contribute to an MR unsaturation even if there is a dramatic change in the carrier balance and mobility, etc.

**Unique electronic structure of WTe$_2$ and its possible relations with the LMR**

LMR is an unusual phenomenon, especially the unsaturated LMR up to 60 T, which has attracted a great deal of attention in condensed matter physics and material science. In order to explore the origin of the LMR in WTe$_2$, high resolution ARPES was used to obtain the electronic structure. The experimental configuration used can be found in methods and the results obtained are shown in **Figure 4**. A series of Fermi pockets was observed in the mapping of the Fermi surface as summarized in **Figure 4(a)**. For each of the cuts marked in **Figure 4(a)**, the photoemission intensity mappings are shown in **Figure 4(b-e)**. The corresponding momentum distribution curves (MDCs) are shown in **Figure 4(f-i)**. Interestingly, each of the bands across the Fermi surface exhibit unique quasi-linear dispersions. By connecting the MDC peaks using dashed lines, several quasi-straight lines were obtained. In addition, all the bands have rather large Fermi velocities, from $1.7 \times 10^5$ to $2.1 \times 10^5$ m/s. Such electronic structures can be compared with the calculations at ambient pressure. Furthermore, by verifying the calculated electronic structures, we found that the quasi-linear dispersion persists at high pressures.

LMR has been reported in several metal/semi-metal single crystal materials, among which Cd$_3$As$_2$[21], TlBiSSe[22] and Na$_3$Bi[23] are Dirac semi-metals and TaAs[24] and NbP[25] are believed to be Weyl semi-metals. BaFe$_2$As$_2$[26] exhibits some Dirac-cone-like structures. Quasilinear band structure are found in almost all these materials with reported LMR besides Bi$_2$Te$_3$[27] and Silver Chalcogenides[28]. Intense spin-orbital coupling and large *g* factors are also believed to contribute to LMR. Strong spin-orbital coupling has recently been confirmed in WTe$_2$ following the observation of circular dichroism[12]. We therefore attribute the unsaturated LMR in WTe$_2$ to its unique electronic structure.

**In summary,** LMR has been demonstrated in WTe$_2$ crystals at high magnetic field and low temperature. It remained robust and became dominant at a pressure of 6.2 GPa. It remained unsaturated up to a large field of 60 T, even when the parabolic semiclassical MR faded out due to broken carrier balance. This suggests MR unsaturation can be universal in WTe$_2$, which in turn suggests its potential application in spintronic devices and high-field sensors.

**Material and Methods**

WTe$_2$ single crystals were grown using the TeBr$_4$ transport method and characterized by X-ray diffraction, as described elsewhere[9]. The high field MR measurements were carried out using a pulse 60 T magnet at Wuhan National High Magnetic Field Center. Sweeps at both negative and positive field polarities were measured for all temperatures. For data collection, SR830 digital lock-in amplifiers and SR560 preamplifiers (Stanford Research Systems) were used. The data were then "symmetrized". The measurements of transport under pressure were carried out using a home-made system in the Hefei High Magnetic Field Laboratory. Its diamond anvil cell has been described elsewhere[9]. Daphne 7373 oil was employed as the pressure-transmitting medium. The ambient-pressure electrical transport measurements were carried out in a Cryomagnetics cryostat in Nanjing University with SR830 and SR850 digital lock-in amplifiers. Ohmic contacts were made using gold wires and silver paste. ARPES experiments were performed at the I05 beamline of Diamond Light Source (DLS) equipped with a Scienta R4000 electron analyzer. The angular resolution was 0.3° and the overall energy resolution was better than 10 meV. The samples were cleaved in situ along the (001) plane and measured under ultra-high vacuum below $5 \times 10^{-10}$ Torr.

## Acknowledgements

The authors would like to thank the National Key Projects for Basic Research in China (Grant Nos. 2013CB922103, 2011CB922103, 2015CB921202 and 2014CB921103), the National Natural Science Foundation of China (Grant Nos. 91421109, 11023002, 11134005, 61176088, 51372112 and 2117109), the NSF of Jiangsu Province (Grant Nos. BK20130054 and BC2013118), the PAPD project, and the Fundamental Research Funds for the Central Universities, for financially supporting this work. The technical assistance provided by Prof. Li Pi and Mingliang Tian of the Hefei High Field Center is also gratefully acknowledged. The insightful discussions with Prof. Haihu Wen from Nanjing University were also extremely helpful. We thank the help from Dr. P. Dudin, Dr. T. Kim and Dr. M. Hoesch at DLS.

**Reference and Notes**

[1] A. Geim, I. Grigorieva, Nature 2013, 499, 419.

[2] G. Fiori, F. Bonaccorso, G. Iannaccone, T. Palacios, D. Neumaier, A. Seabaugh, S. K. Banerjee, L. Colombo, Nature nanotechnology 2014, 9, 768; F. Koppens, T. Mueller, P. Avouris, A. Ferrari, M. Vitiello, M. Polini, Nature nanotechnology 2014, 9, 780.

[3] W. J. Yu, Y. Liu, H. Zhou, A. Yin, Z. Li, Y. Huang, X. Duan, Nature nanotechnology 2013, 8, 952.

[4] L. Britnell, R. Gorbachev, R. Jalil, B. Belle, F. Schedin, A. Mishchenko, T. Georgiou, M. Katsnelson, L. Eaves, S. Morozov, Science 2012, 335, 947.

[5] F. Withers, D. Pozo-Zamudio, A. Mishchenko, A. Rooney, A. Gholinia, K. Watanabe, T. Taniguchi, S. Haigh, A. Geim, A. Tartakovskii, arXiv preprint arXiv:1412.7621 2014.

[6] M. N. Ali, J. Xiong, S. Flynn, J. Tao, Q. D. Gibson, L. M. Schoop, T. Liang, N. Haldolaarachchige, M. Hirschberger, N. Ong, Nature 2014, 514, 205.

[7] I. Pletikosić, M. N. Ali, A. Fedorov, R. Cava, T. Valla, Physical review letters 2014, 113, 216601.

[8] P. Cai, J. Hu, L. He, J. Pan, X. Hong, Z. Zhang, J. Zhang, J. Wei, Z. Mao, S. Li, arXiv preprint arXiv:1412.8298 2014.

[9] X.-C. Pan, X. Chen, H. Liu, Y. Feng, F. Song, X. Wan, Y. Zhou, Z. Chi, Z. Yang, B. Wang, arXiv preprint arXiv:1501.07394 2015.

[10] D. Kang, Y. Zhou, W. Yi, C. Yang, J. Guo, Y. Shi, S. Zhang, Z. Wang, C. Zhang, S.

Jiang, A. Li, K. Yang, Q. Wu, G. Zhang, L. Sun, Z. Zhao, arXiv preprint arXiv:1502.00493 2015.

[11] J. Ye, Y. Zhang, R. Akashi, M. Bahramy, R. Arita, Y. Iwasa, Science 2012, 338, 1193.

[12] J. Jiang, F. Tang, X. Pan, H. Liu, X. Niu, Y. Wang, D. Xu, H. Yang, B. Xie, F. Song, arXiv preprint arXiv:1503.01422 2015.

[13] J. Antoszewski, L. Faraone, OPTOELECTRONICS REVIEW 2004, 12, 347.

[14] J. Antoszewski, G. Umana-Membreno, L. Faraone, Journal of electronic materials 2012, 41, 2816.

[15] J. McClure, Physical Review 1958, 112, 715.

[16] K. K. Huynh, Y. Tanabe, T. Urata, S. Heguri, K. Tanigaki, T. Kida, M. Hagiwara, New Journal of Physics 2014, 16, 093062.

[17] K. Huynh, Y. Tanabe, T. Urata, H. Oguro, S. Heguri, K. Watanabe, K. Tanigaki, Physical Review B 2014, 90, 144516.

[18] Z. Zhu, X. Lin, J. Liu, B. Fauque, Q. Tao, C. Yang, Y. Shi, K. Behnia, Physical review letters 2015, 144, 176601.

[19] At higher temperature, the mobility is strongly suppressed by electron-phonon interaction. The parameters obtained from the mobility spectra at higher temperatures are not exact.

[20] H. Lv, W. Lu, D. Shao, Y. Liu, S. Tan, Y. Sun, arXiv preprint arXiv:1412.8335 2014.

[21] T. Liang, Q. Gibson, M. N. Ali, M. Liu, R. Cava, N. Ong, Nature materials 2014;

A. Narayanan, M. Watson, S. Blake, Y. Chen, D. Prabhakaran, B. Yan, N. Bruyant, L. Drigo, I. Mazin, C. Felser, arXiv preprint arXiv:1412.4105 2014.

[22] M. Novak, S. Sasaki, K. Segawa, Y. Ando, Physical Review B 2015, 91, 041203.

[23] J. Xiong, S. Kushwaha, J. Krizan, T. Liang, R. Cava, N. Ong, arXiv preprint arXiv:1502.06266 2015.

[24] C. Zhang, Z. Yuan, S. Xu, Z. Lin, B. Tong, M. Z. Hasan, J. Wang, C. Zhang, S. Jia, arXiv preprint arXiv:1502.00251 2015.

[25] C. Shekhar, A. K. Nayak, Y. Sun, M. Schmidt, M. Nicklas, I. Leermakers, U. Zeitler, W. Schnelle, J. Grin, C. Felser, arXiv preprint arXiv:1502.04361 2015.

[26] K. K. Huynh, Y. Tanabe, K. Tanigaki, Physical review letters 2011, 106, 217004.

[27] D.-X. Qu, Y. Hor, J. Xiong, R. Cava, N. Ong, Science 2010, 329, 821.

[28] R. Xu, A. Husmann, T. Rosenbaum, M.-L. Saboungi, J. Enderby, P. Littlewood, Nature 1997, 390, 57.

**Figure captions**

**Figure 1. Transport measurements for a WTe$_2$ single crystal. a,** Schematic crystal structure of WTe$_2$. **b,** Temperature dependence of electrical resistivity. Inset: High-resolution TEM image of the crystal and its Fourier transformation. **c,** Upper panel: The magnetoresistance with H ∥ c up to 9 T at temperatures as shown. Inset: Photograph of a single crystal. Lower panel: The field-dependent Hall resistivity $\rho_{yx}$. Inset: Hall coefficient versus temperature at magnetic fields as shown.

**Figure 2. Evidence of carrier balance and linear magnetoresistance (LMR) for a large field. a,** The normalized conductivities calculated from the experimental data at 4.2 K. The blue circles and dashed lines are the experimental data and their analytical representative curves, respectively. The solid lines show the partially normalized conductivities of electron-like (violet) and hole-like (green) carriers obtained from the Kronig-Kramer transformation. **b,** Mobility (μ) spectra of electron-like (left) and hole-like (right) carriers at 4.2 (blue) and 35 K (red) showing the normalized carrier density per unit of μ against μ. Note the long tails up to the high mobility regions at 4.2 K. **c,** Temperature dependence of the carrier density extracted from the mobility spectrum. **d,** Temperature dependence of the carrier mobility. **e,** The difference between the electron and hole populations in WTe$_2$, showing a perfect balance at low temperature. **f,** High-field MR measurements up to 60 T at different temperatures. At high field, the curves at low temperatures appear nearly H-linear. **g,** Curves of MR versus H$^2$. The curves at high temperatures are almost linear in H$^2$.

**Figure 3. Linear magnetoresistance under high-pressure. a,** Magnetoresistance at low temperature under ambient pressure, 1.3, 4.3 and 6.2 GPa. Linear MR was observed under 6.2 GPa. **b,** The magnetic field derivatives of MR. **Figure 4. Electronic structure of WTe$_2$ along the Γ-X direction from ARPES measurements.**

**a,** Schematic of the Fermi surface of WTe$_2$. The red circles refer to the hole pockets, while the blue ones refer to the electron pockets. Different cuts are indicated by the black arrows. **b-e,** Photoemission intensity plot along cut 1, cut 2, cut 3 and cut 4, respectively. The hole bands and electron bands are indicated by the red lines as a guide. **f-i,** The corresponding momentum distribution curves of the spectra shown in panel **b-e**, respectively. The blue dashed lines are traced from the MDC peaks to shown the dispersion of the bands.

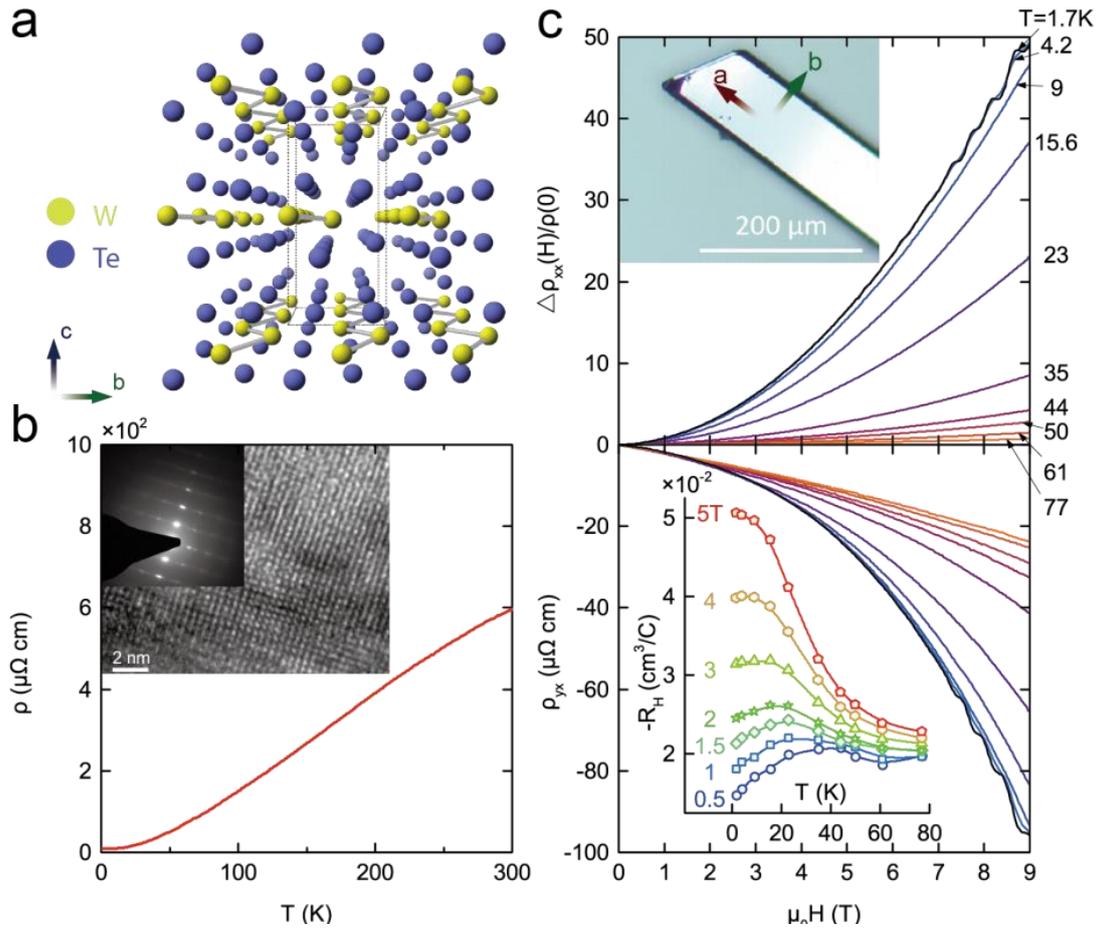

**Figure 1. Transport measurements of WTe$_2$ single crystal.**

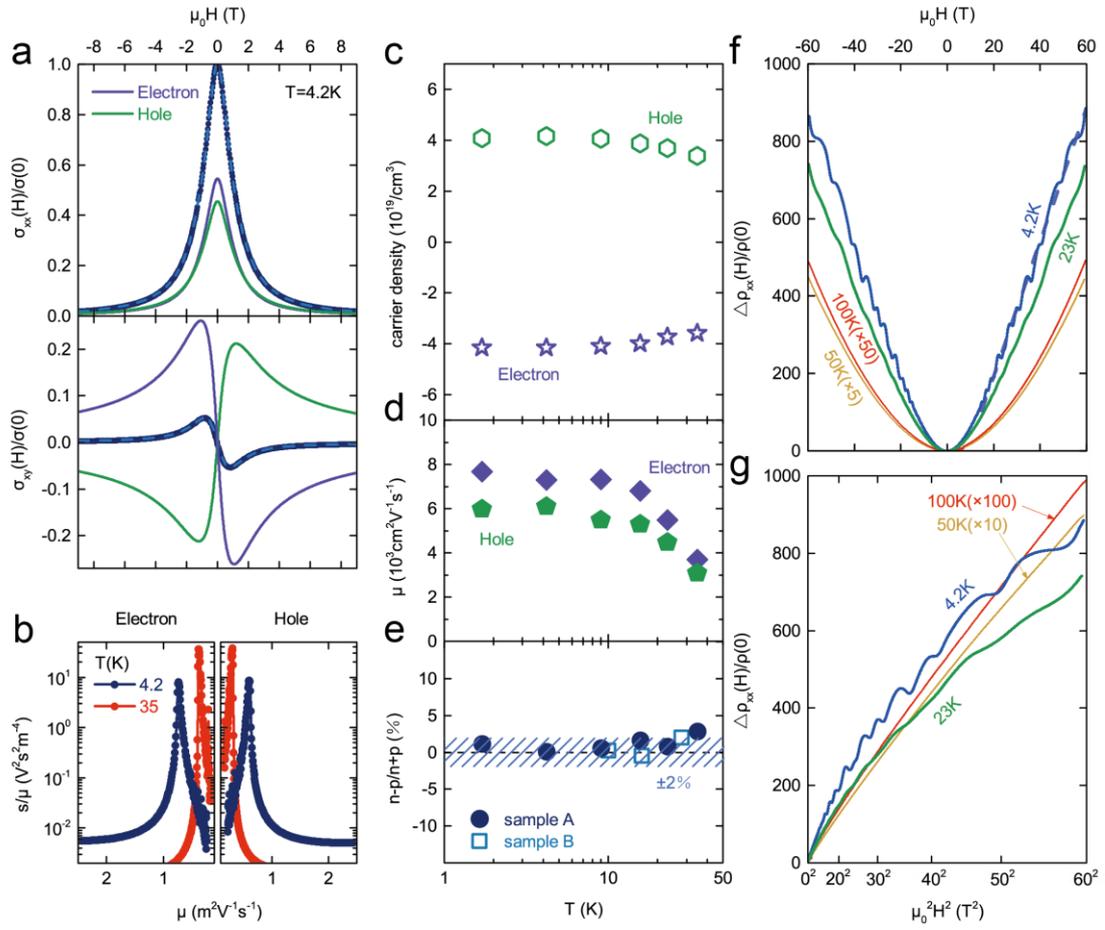

**Figure 2. Evidence of carrier balance and linear magnetoresistance (LMR) in large field.**

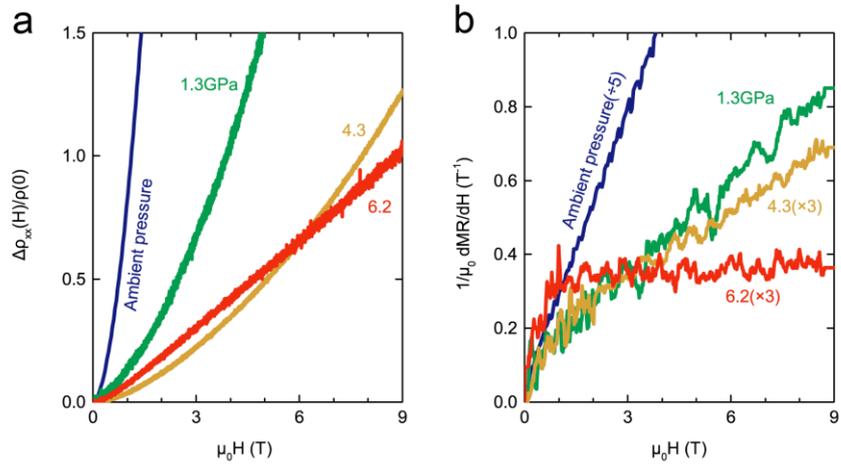

**Figure 3. Linear magnetoresistance under high-pressure.**

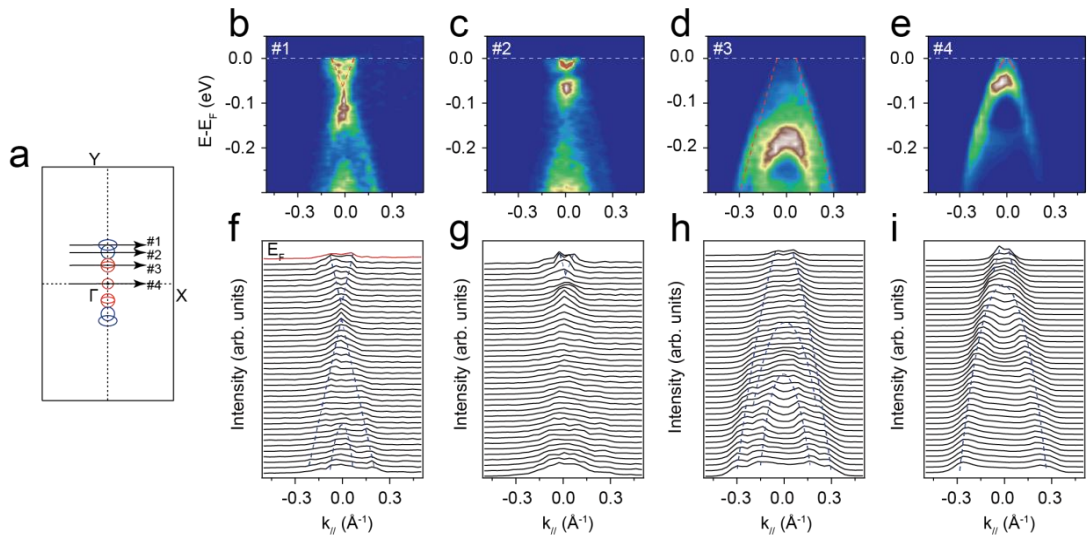

**Figure 4. Electronic structure of WTe$_2$ along the Γ-X direction from ARPES measurements.**

# Supplementary Information for

# Robust linear magnetoresistance in WTe$_2$

Xing-Chen Pan[1], Yiming Pan[1], Juan Jiang[2], Huakun Zuo[3], Huimei Liu[1], Xuliang Chen[4], Zhongxia Wei[1], Shuai Zhang[1], Zhihe Wang[1], Xiangang Wan[1], Zhaorong Yang[4], Donglai Feng[2], Zhengcai Xia[3], Liang Li[3], Fengqi Song[1,*], Baigeng Wang[1,*], Yuheng Zhang[4], Guanghou Wang[1]

**The mobility spectrum analysis**

Traditionally, for a sample involving more than one channel/type of carriers, the conductivity tensor components can be fit as a sum over the m pockets (species) present within the multi-carrier fitting (MCF) system. In this procedure, the carrier densities $n_m$ and corresponding mobilities $\mu_m$ are the fitting parameters, where the number and type of carriers need to be assumed (typically between 2 and 5). The main drawback of the MCF is its arbitrariness. A major advantage of mobility spectrum analysis (MSA) over the MCF procedure is that it is inherently nonarbitrary, i.e., no prior assumption of number or type of carriers is required. The resulting mobility spectra are very useful to analyze the electrical conduction and carrier properties than conventionally obtained concentrations and mobilities.

The conductivity tensor obtained from magnetoresistance and Hall effect data, can be transformed into carrier density as a function of mobility, called the mobility

* Corresponding authors. F.S. (songfengqi@nju.edu.cn) and B.W. (bgwang@nju.edu.cn). Fax:+86-25-83595535

spectra. Instead of assuming the number of electron and hole pockets, the starting point for the MSA is to allow the existence within the sample of a continuous distribution of electron-like and hole-like carriers of any mobility. Therefore, the reduced conductivity tensor is given by

$$\sigma_{xx} = \sigma_{xx}^n + \sigma_{xx}^p = \int \frac{s^{(n)}(\mu)}{1+\mu^2 B^2} d\mu + \int \frac{s^{(p)}(\mu)}{1+\mu^2 B^2} d\mu$$

$$\sigma_{xy} = \sigma_{xy}^n + \sigma_{xy}^p = -\int \frac{s^{(n)}(\mu)\mu B}{1+\mu^2 B^2} d\mu + \int \frac{s^{(p)}(\mu)\mu B}{1+\mu^2 B^2} d\mu$$

where the reduced conductivity are normalized to the conductivity $\sigma_0$ at zero magnetic field. In the preceding, $\sigma^n, \sigma^p$ stands for partial longitudinal and transverse conductivities due to electron-like or hole-like carriers. The various electron and hole pockets then appear as the given mobility peaks in the mobility spectrum. If the electron and hole terms could be separated, the number and the average mobility of each kind of carrier can be found as,

$$K = \frac{\sigma_0}{e} \int \frac{s^{(k)}(\mu)}{\mu} d\mu$$

$$\bar{\mu} = \frac{\sigma_0 \int s^{(k)}(\mu) d\mu}{eK}$$

The parameters K (i.e. n,p) and $\bar{\mu}$ are calculated for a selected range of temperature. The separation of two type partial conductivities can be obtained by the famous Krammers-Kronig relations, yielding

$$\frac{P}{\pi} \int \frac{\sigma_{xx}}{B-B'} dB' = -\sigma_{xy}^n + \sigma_{xy}^p$$

$$\frac{P}{\pi} \int \frac{\sigma_{xy}}{B-B'} dB' = \sigma_{xx}^n - \sigma_{xx}^p$$

Here P denotes the principal part of the integral and the limits of the integrals are from negative infinity to infinity. Practically, in order to carry out the KK transformation analytically, one can fit the real data of conductivities to the linear combinations of Lorentzian components within the MCF. In our case of the experimental data, a linear composed of three Lorentzian components has been used to represent $\sigma_{xx}$, whereas up to five components were necessary to reproduce $\sigma_{xy}$. However, it should be noted that the parameters of the MCF are just a set of analytic representations for the experimental data. The whole algorithm for numerical calculation can be found in the literature cited in the maintext [Ref. 19,20]. To estimate the mobility spectra and avoid the inherent instability in the final spectrum, we discretized the mobility space in a logarithmic equally-spaced grid including up to 500 points. The continuous spectra obtained from the $\sigma_{xx}$ datasets is shown in Figure 2, which is in agreement with that independently obtained from $\sigma_{xy}$.

Finally, we consider the long tail features of mobility spectrum as a signature of the transport of the LMR, i.e. $\sigma_{xx}^L \propto 1/B$. The mobility spectra of the LMR can be deduced by

$$\frac{1}{B} \propto \int \frac{s^L(\mu)}{1+\mu^2 B^2} d\mu$$

From the above equation, one can judge that $s^L(\mu)$ is roughly constant, and then it produced a long tail up to the high mobility region in spectrum $\frac{s^L(\mu)}{\mu}$, as shown in **Figure 2(b)**.